\documentclass[journal]{IEEEtran}

\usepackage{cite}
\usepackage{amsmath,amssymb,amsfonts}
\DeclareMathOperator*{\argmax}{arg\,max}

\usepackage{graphicx}
\usepackage{epsfig,epstopdf} 
\usepackage{tabularx}
\usepackage{textcomp}
\usepackage{mathptmx}
\usepackage{times}
\usepackage{subcaption}
\usepackage[strict]{changepage}
\usepackage[keeplastbox]{flushend}
\usepackage{caption}
\usepackage{subcaption}
\usepackage{comment}
\usepackage{hyperref}
\usepackage[utf8]{inputenc}
\usepackage{authblk}
\usepackage{xcolor}   
\usepackage{multicol}

\usepackage[left=.45in,right=.45in,top=.6in,bottom=.6in,headheight=14.5pt]{geometry}
\usepackage{array,multirow,textcomp}
\usepackage[T1]{fontenc}

\newcolumntype{L}[1]{>{\raggedright\let\newline\\\arraybackslash\hspace{0pt}}m{#1}}

\newcolumntype{C}[1]{>{\centering\let\newline\\\arraybackslash\hspace{0pt}}m{#1}}

\def\BibTeX{{\rm B\kern-.05em{\sc i\kern-.025em b}\kern-.08em
    T\kern-.1667em\lower.7ex\hbox{E}\kern-.125emX}}

\makeatletter
\def\hlinewd#1{%
\noalign{\ifnum0=`}\fi\hrule \@height #1 %
\futurelet\reserved@a\@xhline}
\makeatother

\begin{document}
\markboth{Vol.~1, No.~3, July~2017}{0000000}

\title{Deep Single Shot Musical Instrument Identification using Scalograms}

%
\author{\IEEEauthorblockN{Debdutta Chatterjee\IEEEauthorrefmark{1}, Arindam Dutta\IEEEauthorrefmark{2},
Dibakar Sil\IEEEauthorrefmark{3},Aniruddha Chandra\IEEEauthorrefmark{1}}\\
\IEEEauthorblockA{\IEEEauthorrefmark{1}Department of Electronics and communication Engineering,
National Institute Of Technology , Durgapur,713209, India\\
\IEEEauthorrefmark{2}Department of Computational and Data Sciences, Indian Institute of Science, Bangalore, India\\
\IEEEauthorrefmark{3}Steradian Semiconductors Pvt. Ltd., Bangalore, India..\\
Senior Member, IEEE}%
\thanks{Manuscript received June 07, 2021; revised XXX xx, 2021; accepted XXX xx, 2021. Corresponding author: Aniruddha Chandra.}%
\thanks{D. Chatterjee and A. Chandra are with the ECE Department, National Institute of Technology, Durgapur 713209, WB, India (e-mail: aniruddha.chandra@ieee.org).}%
\thanks{A. Dutta is with the Department of Computational and Data Sciences, Indian Institute of Science, Bangalore, India.}%
\thanks{D. Sil is with Steradian Semiconductors Pvt. Ltd., Bangalore, India.}%
\thanks{This work is supported by Core Research Grant (CRG), Science and Engineering Research Board, Department of Science and Technology, Government of India, Grant No. CRG/2018/000175 and the Research Initiation Grant (RIG), NIT Durgapur, Grant No. 996/2017.}%
}

\markboth{IEEE System Journal,~Vol.~X, No.~x, Xxxxxx~2021 }
{CHATTERJEE \MakeLowercase{\textit{et al.}}: Deep Single Shot Music Instrument Identification using Scalograms}




\maketitle

\begin{abstract}
Musical Instrument Identification has for long had a reputation of being one of the most ill-posed problems in the field of Musical Information Retrieval(MIR). Despite several robust attempts to solve the problem, a timeline spanning over the last five odd decades, the problem remains an open conundrum. In this work, the authors take on a further complex version of the traditional problem statement. They attempt to solve the problem with minimal data available - one audio excerpt per class. We propose to use a convolutional Siamese network and a residual variant of the same to identify musical instruments based on the corresponding scalograms of their audio excerpts. Our experiments and corresponding results obtained on two publicly available datasets validate the superiority of our algorithm by $\approx$ 3\% over the existing synonymous algorithms in present-day literature.  
\end{abstract}


\IEEEpeerreviewmaketitle

\section{Introduction}

\IEEEPARstart{M}{usic} forms a major component of the entertainment industry \cite{rafii2018overview}. Applications in the domain of musical signal processing are widely increasing, which includes automatic music transcription, beat tracking and extraction of melody from music, to name a few \cite{Cano}. Proper identification of instruments is critical in musicology to understand the orchestration pattern and how the presence and combination of different instruments emerged over time \cite{chon2018exploratory}. Further, procedures such as mixing and instrument wise equalization require proper identification of musical instruments. Also, archiving and cataloguing require musical instrument classification \cite{wold1996content}. Traditional machine learning algorithms based on manual feature extraction has been extensively used for musical instrument classification \cite{Han}, \cite{ghosh2018music}. Most of the musical instrument classification work used studio generated music\cite{livshin2004musical}. This music differs from music generated in an orchestra as it consists of more than a single instrument, consists of the superposition of concurrent notes, echoes, and other background noise. One of the primary works in this field was \cite{marques1999study}, in which the authors classified 8 instruments (bagpipes, clarinet, flute, harpsichord, organ, piano, trombone, and violin) using Mel cepstral coefficients as features and implemented the Support Vector Machine classification algorithm with 70\% accuracy. In a later paper \cite{brown2001feature}, authors classified 4 different instruments((flute, sax, oboe, and clarinet) using cepstral coefficients, constant-Q coefficients, spectral centroid, auto-correlation coefficients. They obtained an accuracy of around 80\% by choosing the optimal set of parameters as features. Authors of \cite{livshin2004musical}  classified 7 instruments (bassoon, clarinet, flute, guitar, piano,
cello, and violin) with a recognition accuracy of 88.13\%. They used 62 different features and used the Gradual Descriptor Elimination feature selection method to determine the most suitable features for each class.
This feature engineering requires immense knowledge of a particular domain and is tedious as repetitive and hand-crafted features are sometimes internally redundant.

\begin{figure*}[htb!]
\begin{center}
\includegraphics[width=450pt]{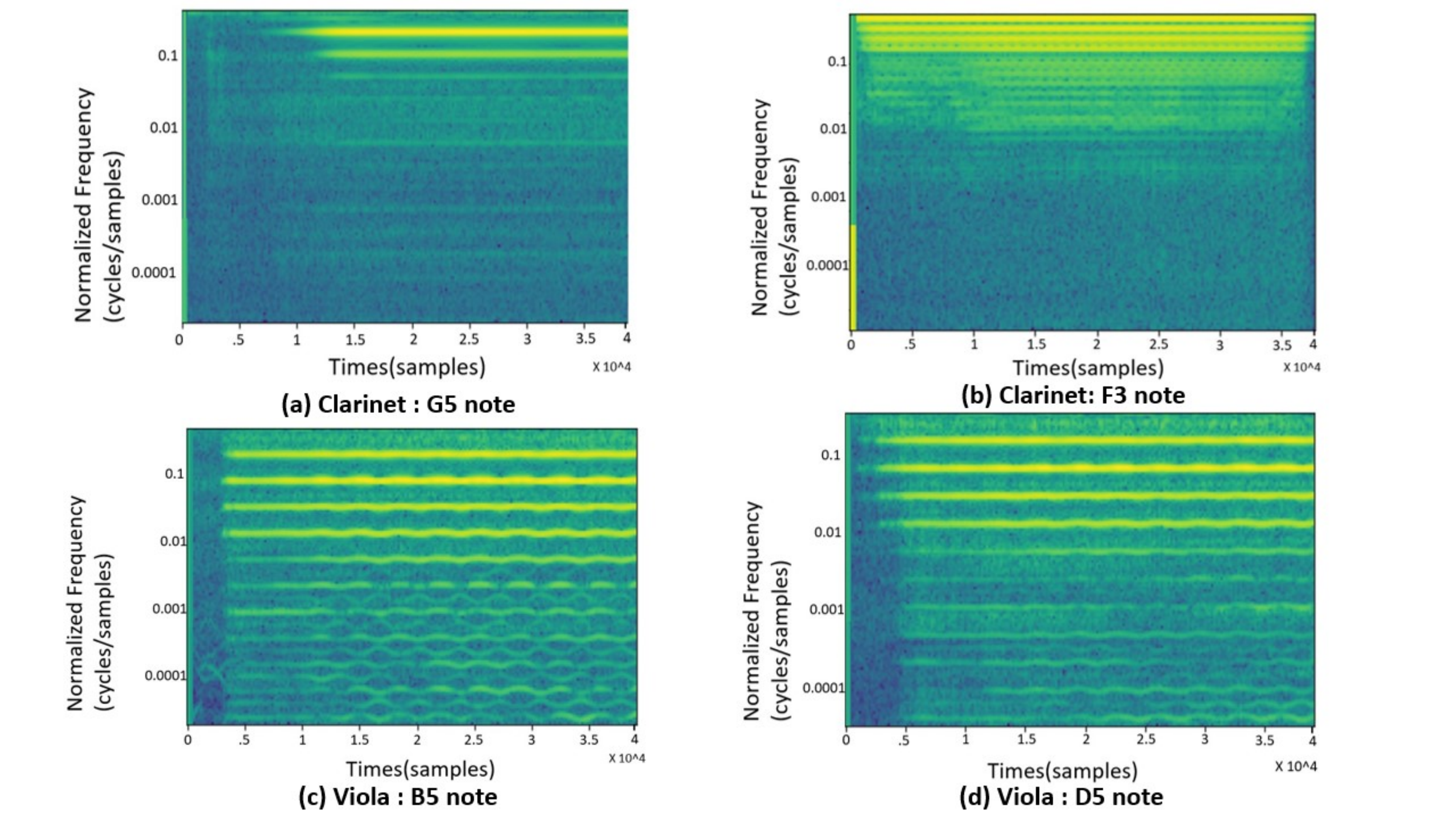}
\caption{Spectograms}
\label{fig:Specto}
\end{center}
\end{figure*}

Existing literature presents some of the popular data-driven classification algorithms, such as musical instrument identification using short-time Fourier transform \cite{Martin}, \cite{Pons} and Mel-frequency, cepstral coefficients (MFCCs) \cite{Bhalke}. Of late, the data-driven model based on convolutional neural networks for automatic musical instrument classification \cite{duttacnn}, and wavelet-based models\cite{Ren} has shown promising results. Authors of \cite{gururani2019attention} present an interesting algorithm based on attention mechanisms of musical instrument classification. Similarly, \cite{rosner2018automatic} presents a study of musical genre classification based on musical instrument identification. Authors of \cite{hung2018frame} study the frame-level classification of musical instruments based on timbre and speech, whereas \cite{taenzer2019investigating} presents a study of musical instrument classification for western classical musical recordings. These recent works demonstrate the importance and associated difficulties of musical instrument classification in the field of music information retrieval.

Data-driven approaches require colossal data sets, but there is a paucity of these resources in this domain as most music databases are copyrighted. 
Thus lack of standard datasets in the field of musical instrument identification necessitates us to venture out to the field of one-shot musical instrument classification, which is an onerous problem that has never been dealt with before. In cases where a quick decision of identifying musical instruments is required, one-shot learning can be efficient. For example, Alexa is playing a song, and a quick check of which instruments are being played in real one-shot learning can be instrumental.

Instead, simple Convolutional Neural Network-based architectures, which are an integral part of the rapidly growing field of Deep Learning \cite{lecun2015deep}, have shown promising results with less parameter complexity \cite{han2016deep}. Recent literature as presented in\cite{duttacnn} shows that scalograms fed to a simple CNN architecture can classify instruments with an accuracy of 85\%. Taking inspiration from the same, we plan to use scalograms to feed into our deep siamese networks for one-shot classification.

It is critical to note that one-shot musical instrument classification techniques are relatively immature and have received limited attention by machine learning for the audio and speech community. The CNN based siamese network as presented in \cite{koch2015siamese} has outperformed several complex algorithms in the single-shot classification of handwritten characters. In this paper, we devise a novel approach to this problem by using deep convolution Siamese neural networks and the residual variant of the same.   
Since the Siamese network takes 2-D vectors as input, musical sounds being a 1-D vector, is converted through the use of short-time Fourier transform to extract joint time-frequency features of an audio signal and later to spectrograms. A spectrogram is a visual representation of the spectrum of frequencies of a signal as it varies with time.

Spectrograms are used extensively in music, sonar, radar, speech processing, seismology, and others. A bank of  band-pass filter can generate a spectrogram by STFT or by wavelet transform, the spectrogram obtained in later cases is called a scalogram. As mentioned in \cite{ren2018deep}, scalograms have several advantages over the spectrogram in terms of windowing function that can be stretched in time and frequency domain, negating the problem of window selection of STFT. Moreover, scalograms have performed well at higher noise levels, making them suitable for the characterization of real-world signals. Thus, a scalogram has been used as equivalent 2D representation is fed to the Siamese network to harness the classification capability of the network.

The major contributions of this work are as follows:
\begin{itemize}
 \item A comparative study of both scalograms and spectrograms as time-frequency representation features for one-shot classification is performed.  

  \item A Robust Siamese network architecture has been proposed to one-shot classify the musical instrument with appreciable accuracy, thus ensuring that we can use this network to obtain appreciable accuracy despite the dearth of data.
  
  \item To reduce the number of parameters and thus the memory footprint of these architectures - we also propose a residual version of the proposed Siamese architecture.
\end{itemize}

The rest of the paper is organized as follows: Section 2 presents a brief overview of the datasets used in our study, section 3 presents a detailed account of the methodology, section 4 presents the experiments performed and corresponding results obtained and finally, section 5 concludes the paper.

\section{Datasets}
The music instrument datasets from Kaggle \cite{dataset1}, and ISMIR \cite{dataset2} are used as they satisfy the basic conditions such as transparency, openness and proper annotations. Both the datasets contain enough musical instruments for testing and training our model. 

We have taken the Morse analytic wavelet transform for generating scalograms. The short-time Fourier transform (STFT) suits better for non-stationary signals, while continuous wavelet transform (CWT) gives high time-frequency resolution and is better suited for analyzing signals that contain non-periodic and fast transients features.
\begin{figure*}[ht]
\begin{center}
\includegraphics[trim = 2cm 0cm 2cm 0cm, clip, scale=0.50]{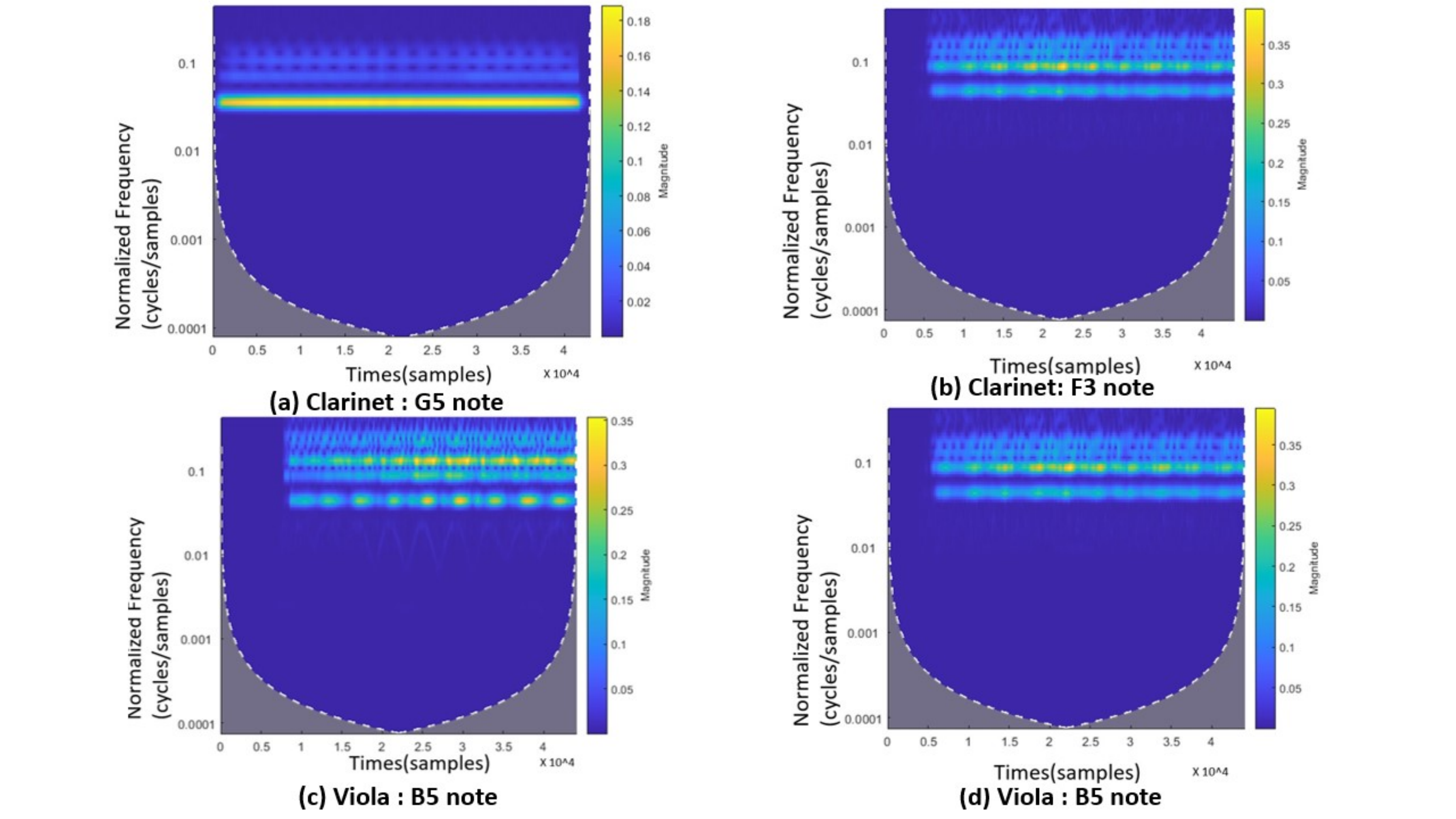}
\caption{Scalograms corresponding to different notes of different instruments showing how different notes of different instruments may have similar scalograms - justifying the importance of the problem.}
\label{fig:Scalogram}
\end{center}
\end{figure*}

\begin{subsection}{Dataset from Kaggle}
The dataset \cite{dataset1} has been procured by recording $14$ different musical instruments for $1$ s at a sampling rate of $44.1$ kHz. Table \ref{Table:1} gives the figures for class and categories of all the instruments recorded.
\begin{table}[ht]
\caption{Instrument samples in Kaggle dataset}
\centering
\begin{tabular}{|p{2.3cm}|p{3.2cm}|p{2cm}|}    \hline
{\bf String} & {\bf Brass} & {\bf Woodwind} \\ \hline
 Bass[Double] (153) & French Horns (166)        & Clarinet (481) \\
 Cello (227)        & Saxophone [Alto] (480)    & Flute (454)    \\
 Guitar (420)       & Saxophone [Soprano] (284) & Oboe (360)     \\
 Viola (220)        & Tuba (560)                & Trombone (312) \\
 Violin (560)       &                           & Trumpet (246)  \\
 \hline
\end{tabular}
\label{Table:1}
\end{table}

Although the dataset contains $2500$ audio signals, only $1540$ audio excerpts were used as an equal amount of data was required for each instrument. Scalograms of size $[224,224,3]$ produced by CWT are fed to the proposed networks for training and validation of one-shot learning.

\begin{figure*}[ht]

  \centering
  \includegraphics[width=0.85\linewidth]{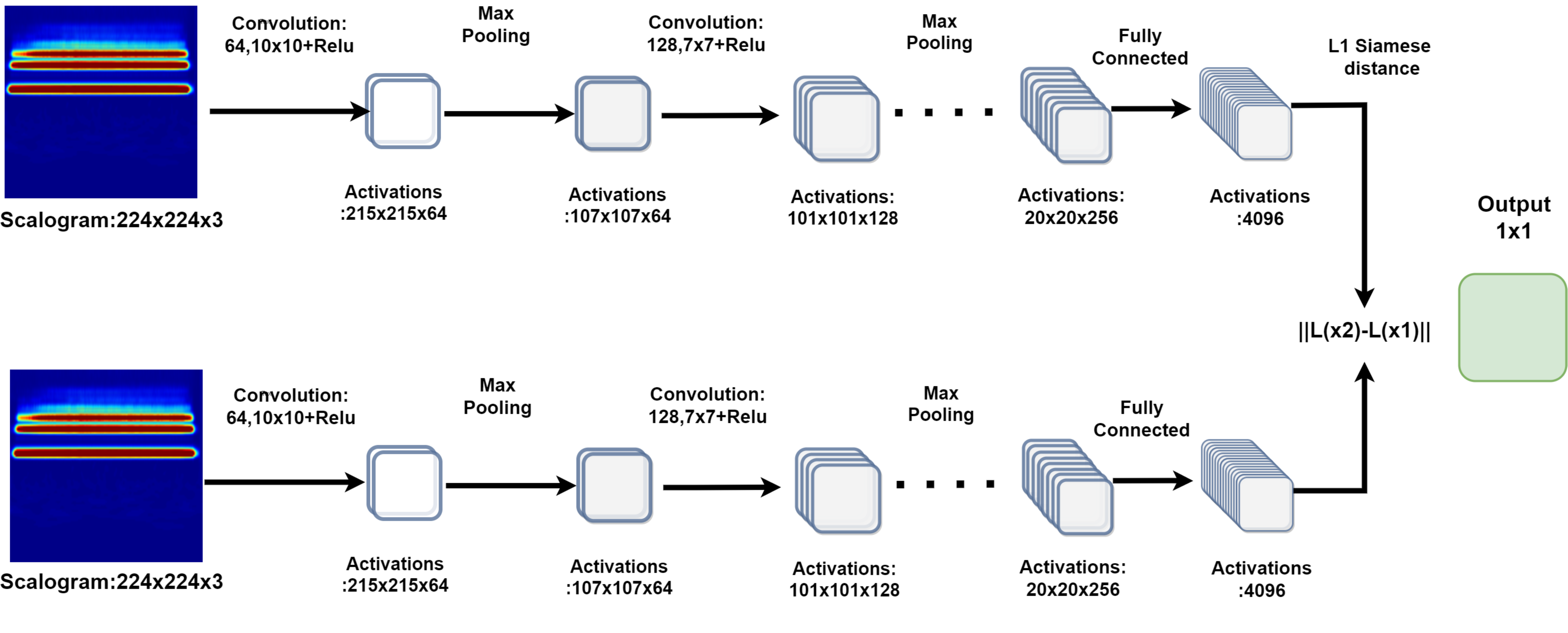}  
  \caption{Schematic of the Siamese neural network architecture for musical instrument identification using scalogram }
  \label{fig:deep-Siamese_schematic}
\end{figure*}

It should be taken into account that the scalogram of the same instrument(clarinet) at two different notes has a significant amount of difference. In contrast, different instruments of similar classes (violin, viola) have a striking similarity. This makes the one-shot classification task unfrivolous.
\end{subsection}

\begin{subsection}{Dataset from ISMIR}
\subsubsection{Dataset from ISMIR}
The second dataset used for our work is the open-source ISMIR dataset \cite{dataset2} which consists of $3$ s long clips of various musical instruments sampled at $44.1$ kHz. As shown in Table \ref{Table:2}, the dataset consists $6715$ number of audio data of $10$ musical instruments. In order to maintain consistency, $388$ samples of each category of instruments (total of $3880$) were considered for training.

\begin{table}[ht]
\caption{Instrument samples in ISMIR dataset}
\centering
\begin{tabular}{|p{3cm}|p{2.5cm}|p{2cm}|}     \hline
 {\bf String} & {\bf Brass} & {\bf Woodwind} \\ \hline
 Piano (721)           & Saxophone (626) & Clarinet (505) \\
 Cello (388)           & Trumpet (577)   & Flute (451)    \\
 Electric Guitar (760) &                 & Organ (682)    \\
 Violin (580)          &                 &                \\
 Acoustic Guitar (637) &                 &                \\
 \hline
\end{tabular}
\label{Table:2}
\end{table}

A significant difference of the second dataset from the first one is, apart from the signal of the primary instrument, each audio clip also contains notes from other instruments or human voices in the background. 

Similar to the previous dataset, scalograms are obtained for each audio data of size $[656,875,3]$. We further resize these images to size $[224,224,3]$ to obtain the final dataset fed to the model.
It should be noted that although the audio data for a particular instrument consist of primarily that instrument but also contains notes from other instruments or human voices in the background. 
Similar to the previous dataset, scalograms are obtained for each audio data of size $[656,875,3]$. We further resize these images to size $[224,224,3]$ to obtain the final dataset fed to the model.

\end{subsection}

\section{Methodology}

\begin{subsection}{Problem Statement}
Given the set of audio excerpts $\mathbf{X}$, such that $\mathbf{x_{1}}$, $\mathbf{x_{2}}$, .... , $\mathbf{x_{k}}$ with corresponding elements $x_{11}$, $x_{12}$, .... , $x_{1a}$; $x_{21}$, $x_{22}$, .... , $x_{2b}$; $x_{k1}$, $x_{k2}$, .... , $x_{kn}$, are disjoint subsets of $\mathbf{X}$ with respective correspondence to the labels $y_{1}$, $y_{2}$, .... , $y_{k}$, we wish to find a many-to-one mapping $\mathcal{F}(.)$ such that, \begin{equation}
    y_{i}\mathbf{1}(i=j) = \mathcal{F}(x_{ip}, x_{jq}; \Theta)
\end{equation}  
where $i,j \leq k$ and $y_{i}$ is the true label corresponding to some $x_{ip}$ $\in$ $\mathbf{x_{i}}$ $\in$ $\mathbf{X}$ and $\Theta$ representing the parameters of the network $\mathcal{F}(.)$. $\mathbf{1(.)}$ is the Indicator Function.
\end{subsection}
\begin{subsection}{Short Time Fourier Transform vs Continuous Wavelet Transform}

Though the importance of STFT is well known in signal processing society, STFT suffers from a major drawback of window length selection and gives shallow time-frequency resolution.
The Short-time Fourier transform (STFT) suits better for non-stationary signals. 

On the other hand, Continuous Wavelet transform is better suited for analyzing signals that contain non-periodic and fast transients features.

\begin{figure}
    \includegraphics[width=1\linewidth]{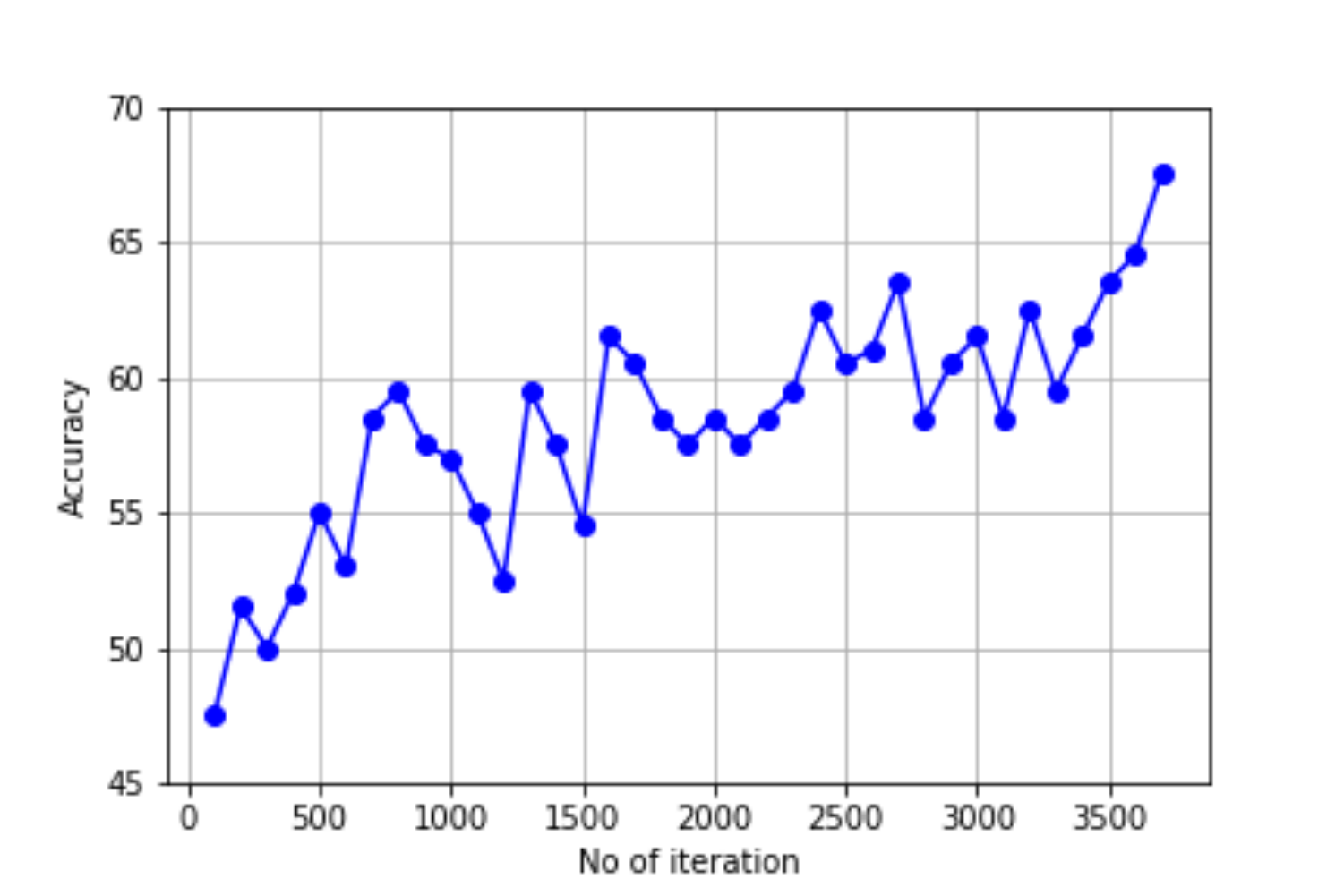}
    \caption{Accuracy of training set vs Epoch}
    \label{Fig:Accuracy_of training set vs Epoch}
\end{figure}

Wavelet basis is beneficial for representing a signal in terms of domain basis and their coefficients. Unlike the Fourier transform (FT) or the Short-Time Fourier Transform (STFT), the CWT analyzes a signal at different frequencies with different resolutions. Wavelets are well localized, and very few coefficients are needed to represent the signal.
In our case, we have taken the Morse analytic wavelet transform.

\end{subsection}

\subsection {Proposed Network Architectures and One Shot Learning Algorithm}

Taking inspiration from \cite{koch2015siamese}, we base our architecture on deep Siamese networks. Siamese (twin) networks are nothing but congruent networks that are tied by the same weights. When two networks share weights, it is obvious to expect that when the same image will be passed through both the networks, the corresponding feature maps and hence the single-dimensional feature vectors obtained at the penultimate layer will be similar. 
VGG  type architecture has been used, consisting of just 14 layers.  The network consists of Convolution Filters of size (10x10)
progressively reduced by the size of 3.
To maximise efficiency, the number of Convolution filters is kept as a multiple of 16.
Each convolution Block is followed by a Relu and Max-Pooling layer of size (2x2)
The feature vector of the final Convolution Block is flattened into a single vector of length 4096.
Figure \ref{fig:deep-Siamese_schematic} demonstrate the network architecture in  detail.

We have used the difference between the feature vectors as a weighted L1 distance in the last layer so that the last layer is sparse enough for easy processing. Further, we use the sigmoid function to squash the values of the elements of the last layer to [0,1] and use it as a probabilistic measure. The binary cross-entropy objective has been used for training the model, and the Loss-Epoch training plot has been shown in figure \ref{Fig:Accuracy_of training set vs Epoch}.

Mathematically, Prediction Vector (P) is given by equation \ref{predvec}.
\begin{equation}
     P = \sigma(\sum_{j}\gamma_{j}|M_{1,L-1}^{j} - M_{2,L-1}^{j}|)
     \label{predvec}
\end{equation}
Here, $\sigma(.)$ is the sigmoid function and $\gamma_{j}$ are additional weights (parameters) of the network which are duly learned during training.
$M_{1}(.)$ and $M_{2}(.)$ are the two component networks of the Siamese network.
\\
The binary cross entropy loss function can be expressed by equation \ref{bce}.
\begin{equation}
\begin{split}
    \mathbf{L}(x_{1},x_{2}) & = \mathbf{p}(x_{1},x_{2})\log(\mathbf{p}(x_{1},x_{2})) + \\
    & (1 - \mathbf{p}(x_{1},x_{2}))\log(1 - \mathbf{p}(x_{1},x_{2}))
    \label{bce}
    \end{split}
\end{equation}
Here, $\mathbf{p}(x_{1},x_{2})$ = 1, if $x_{1}$ and $x_{2}$ belong to the same audio instrument class, otherwise 0. 

The network takes input in the form of a pair of scalograms (or spectrograms), each of which has dimensions [224x224x3]. The number of parameters of the network stands out at $\approx$ 420 million (420,646,209, to be exact). Thus, the network can overfit to a large extent. This necessitates us to use pairwise training and dropout techniques \cite{srivastava2014dropout}, which make the network robust. The use of max-Pooling layers after convolutional layers ensure dimensionality reduction. This makes the network focus far smaller subspace of the input data, thus aiding in increased classification accuracy and lower memory requirements. The use of ReLU non-linearity \cite{nair2010rectified} ensures that the activations do not die out in deeper layers of the network. The extremely popular optimizer, Adam \cite{kingma2014adam} has been used with a constant learning rate of \(6 \times 10^{-4}\) and default hyper-parameter values to obtain easy convergence.

Making networks go deeper often makes them overfit on the training set, resulting in poor performance on the test set. Residual networks, powered by skip connections as shown by equation \ref{skip-c}, initially presented in \cite{he2016deep} have revolutionized the idea of deeper networks with a lesser number of parameters. These networks also mitigate the problem of vanishing gradients, a problem extremely common to deep CNNs. Taking inspiration from the same, we modify the initially presented architecture to contain residual connections. We find that the number of parameters drops by a factor of $\approx$ 18$ \times $ to $\approx$ 23 million (23,553,025, to be exact), without any sacrifice inaccuracy. As earlier, Adam optimizer is have been used with a learning rate of \(5 \times 10^{-4}\) and other hyper-parameters set to default values. Mathematically,
for some part of a traditional deep CNN, say $x$ is the input and $\mathcal{H}(x)$ = $\mathcal{F}(x)$ is the output. Incorporating skip connection, the input-output relation can be given by equation \ref{skip-c}.
\begin{equation}
\mathcal{H}(x) = \mathcal{F}(x) + x
    \label{skip-c}
\end{equation}

The procedure of learning adequate features from a small data set can be very computationally expensive, and such a problem can prove very daunting. One-shot learning is one such problem in which predictions are made based on a single example. 
Once the network has been optimally trained, we are all set to test and demonstrate the discriminative potential of the network, not just on new data but on data from an unknown distribution.
Given a query scalogram $x_{q}$ and corresponding scalograms $X_{k =1}^{k=C}$ belonging to one of the $\mathbf{C}$ classes, we predict the class $C^{*}$ in accordance to equation \ref{siam-test}.
\begin{equation}
    C^{*} = \argmax_{X^{k}}\mathbf{P}_{k}
    \label{siam-test}
\end{equation}
Here. $\mathbf{P}_{k}$ is the prediction vector from equation \ref{predvec}.

\section{Experiments and Discussion}

The codes for the project were executed on two separate systems for two separate tasks. Codes pertaining to CWT of the audio excerpts were written and executed in MATLAB 2019b on a system with 64GB RAM and an Intel-i7 core processor.
\begin{table*}[htb]
\centering
\caption{Accuracy with Kaggle dataset}

\begin{tabular}{|c|c|c|c|c|c|c|c|c|c|}
\hline
{\bf Training} & {\bf Testing} & \multicolumn{4}{c|}{{\bf convolutional Siamese}} & \multicolumn{4}{c|}{{\bf residual Siamese}} \\ \cline{3-10}
{\bf data} & {\bf data} & \multicolumn{2}{c|}{{\bf Scalogram}} & \multicolumn{2}{c|}{{\bf Spectogram}} & \multicolumn{2}{c|}{{\bf Scalogram}} & \multicolumn{2}{c|}{{\bf Spectogram}} \\ \cline{3-10}
{\bf sets} & {\bf sets} & {\bf max} & {\bf mean} & {\bf max} & {\bf mean} & {\bf max} & {\bf mean} & {\bf max} & {\bf mean}\\
 \hline
 2 & 12 & 82\% & 65\% & 74\% & 61\% & 74\% & 60\% & 71\% & 52\%  \\
 5 &  9 & 82\% & 69\% & 76\% & 63\% & 84\% & 70\% & 79\% & 64\% \\
 8 &  6 & 86\% & 74\% & 78\% & 66\% & 90\% & 73\% & 83\% & 68\% \\
10 &  4 & 86\% & 75\% & 80\% & 69\% & 94\% & 81\% & 89\% & 78\% \\
12 &  2 & 92\% & 78\% & 82\% & 73\% & 96\% & 94\% & 89\% & 84\% \\
\hline
\end{tabular}

\label{Table:3}
\end{table*}

\captionsetup{justification=centering}
\begin{table*}[htb]
\caption{Accuracy with ISMIR dataset}
\centering
\begin{tabular}{|c|c|c|c|c|c|c|c|c|c|}
\hline
{\bf Training} & {\bf Testing} & \multicolumn{4}{c|}{{\bf convolutional Siamese}} & \multicolumn{4}{c|}{{\bf residual Siamese}} \\ \cline{3-10}
{\bf data} & {\bf data} & \multicolumn{2}{c|}{{\bf Scalogram}} & \multicolumn{2}{c|}{{\bf Spectogram}} & \multicolumn{2}{c|}{{\bf Scalogram}} & \multicolumn{2}{c|}{{\bf Spectogram}} \\ \cline{3-10}
{\bf sets} & {\bf sets} & {\bf max} & {\bf mean} & {\bf max} & {\bf mean} & {\bf max} & {\bf mean} & {\bf max} & {\bf mean}\\
 \hline
 3 & 8 & 78\% & 55\% & 73\% & 53\% & 74\% & 60\% & 71\% & 52\% \\
 5 & 6 & 82\% & 60\% & 76\% & 59\% & 78\% & 61\% & 73\% & 54\% \\
 7 & 4 & 74\% & 63\% & 72\% & 61\% & 80\% & 61\% & 72\% & 61\% \\
 9 & 2 & 78\% & 61\% & 76\% & 64\% & 82\% & 64\% & 78\% & 63\% \\
\hline
\end{tabular}
\label{Table:4}
\end{table*}

Codes pertaining to the network training and testing were written in the TensorFlow 2.0 environment on the Google Colaboratory and were executed on the 12 GB Tesla K80 GPU. The codes have also been made available for enthusiastic users on \href{https://github.com/Dibakar1/Deep-Single-Shot-Musical-Instrument-IdentificationUsing-Time-Frequency-Localized-Features}{\underline{this}} open-source repository.

\subsection{Dataset from kaggle}
To access and hence justify the potential of our algorithm, we train the network on randomly chosen training sets which are essentially subsets of the dataset. We randomly choose 2, 5, 8,10, and 12 sets of training examples and test the network on the rest of the unknown audio examples. The network is trained and tested ten times with a given number of training and testing sets, each time while testing the maximum and mean accuracy is noted. The final published accuracy is the maximum and means of these ten iterations.
Table \ref{Table:3} shows the accuracy obtained on the Kaggle dataset upon using the two proposed networks, namely the convolutional Siamese network and residual-convolutional Siamese network.

\begin{figure}

\centering 
  \includegraphics[width=1\linewidth]{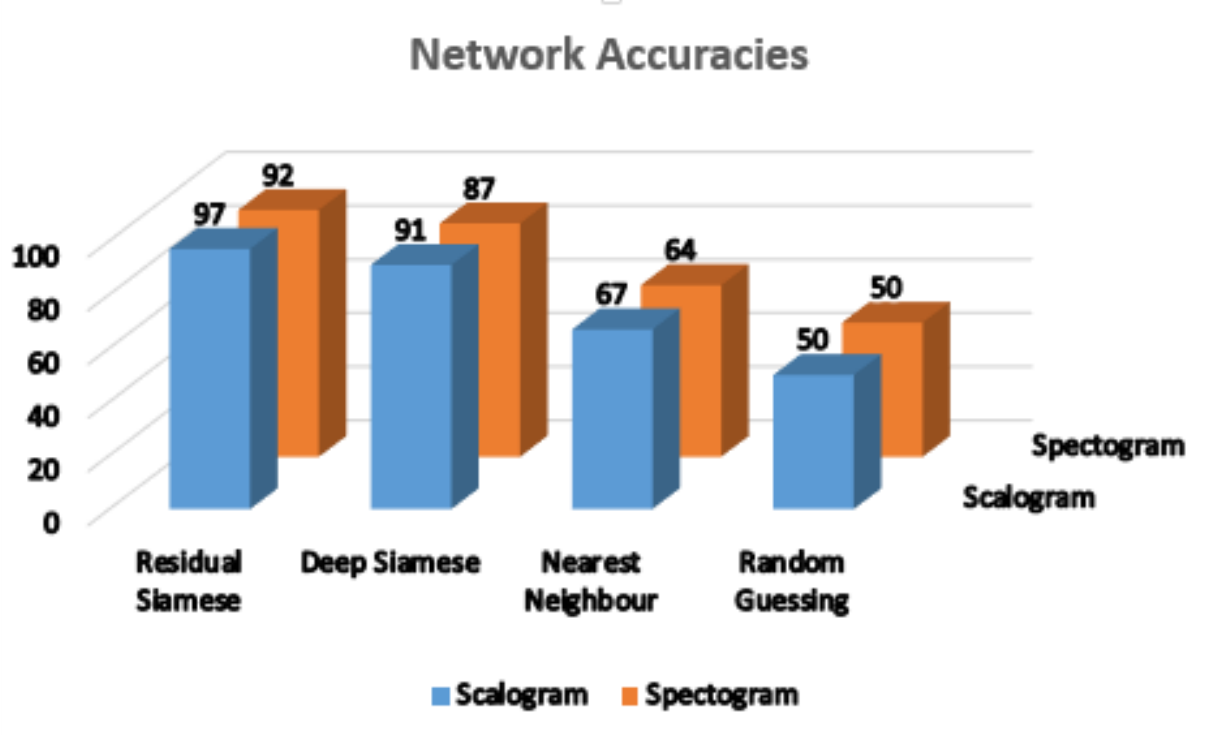}  
  \caption{Comparing best one-shot accuracy from each type of
network against baselines.}
  \label{fig:network-accuracy}
\end{figure}

Even with just two training sets, we have a mean accuracy of around 65 \%, which increases to more than 90 \%, when trained on 12 training sets. The catch here is that most traditional classification algorithms or even conventional DL models as like \cite{duttacnn} or \cite{solanki2019music} would predict this accurately only when the network has had access to scalograms pertaining to all instrument classes. Also, it is obvious that original datasets will have some noise in their audio excerpts, and despite that, the network performs remarkably well. Also, since the dataset was not manually screened for ill-audio excerpts, which would have resulted in higher accuracy, the difference in best accuracy and mean accuracy is just obvious. Figure \ref{fig:network-accuracy}. sums up a comparative study against synonymous baselines.

\subsection{ISMIR dataset}

The ISMIR dataset is one of the few standard datasets in this field for musical instrument classification and hence has been studied thoroughly. What we observe is that our proposed networks, albeit they do not predict instrument classes more accurately than convention DL models as like \cite{solanki2019music}, performs pretty well. Tables \ref{Table:4} show the performance of the networks under varied conditions of training and testing. 
 While best accuracy hovers around 80\%, mean accuracy more often than comes out to be around 65\%. These numbers are justifiably appreciable. Thus, our proposed networks can quite efficiently categorize a musical instrument from its audio excerpt.


\section{Conclusion}
In this work, we propose a novel single-shot musical instrument recognition algorithm. Given that properly annotated data for musical instruments is not cornucopious, our proposed algorithm, with its' abundant pragmatism, fits right into the gap. We base our algorithm on Siamese convolutional networks, which in effect study the similarities of two given scalograms rather than memorizing the feature spaces corresponding to scalograms of one particular musical instrument. Our experiments show that we can achieve state-of-the-art results even with just one audio excerpt example per class. However, our network suffers a major drawback in terms of network parameters, which albeit are low in terms of memory utilization on GPUs but are not suited for portable device applications. In our future works, we will try to develop an online-learning method based on light-weight Convolutional or Recurrent Neural architectures, which would make this algorithm a perfect match with low-power portable devices.

\bibliographystyle{IEEEtran}
\bibliography{bilbiography.bib}

\end{document}